\documentstyle[twocolumn,epsf]{jpsj}
\def\runtitle{ 
Magnetic Fluctuations in a Charge-Ordered State  
of the One-Dimensional Extended Hubbard Model with  
a Half-Filled Band

}
\def\runauthor{Hideo {\sc Yoshioka}}
\def\d{{\rm d}}

\def\e{{\rm e}}

\def\ek{\epsilon_{k}}
\def\ekq{\epsilon_{k+q}}
\def\Ek{E_{k}}
\def\Ekq{E_{k+q}}

\def\ggs{\buildrel\textstyle > \over {\hbox{\raise0.2ex\hbox{$\sim$}}}}
\def\lls{\buildrel\textstyle < \over {\hbox{\raise0.2ex\hbox{$\sim$}}}}
\def\gsim{\,\lower0.75ex\hbox{$\ggs$}\,}
\def\lsim{\,\lower0.75ex\hbox{$\lls$}\,}

\def\im{{\rm i}}
\def\ie{{\it i.e.}, }

\def\deltau{\partial_\tau}
\def\on{\omega_n}
\def\jo #1#2#3#4{#1 {\bf #2} (#3) #4}   

\def\PRB{Phys.\ Rev.\ B}
\def\PRL{Phys.\ Rev.\ Lett.}

\def\JPSJ{J.\ Phys.\ Soc.\ Jpn.}

\def\SM{Synth.\ Met.}

\def\BCSJ{Bull.\ Chem.\ Soc.\ Jpn.}
\def\JPCS{J.\ Phys.\ Chem.\ Solids}
\title
{
Magnetic Fluctuations in a Charge-Ordered State  
of the One-Dimensional Extended Hubbard Model with  
a Half-Filled Band
}

\author{
Hideo {\sc Yoshioka}\footnote{E-mail: yoshioka@phys.nara-wu.ac.jp} 
}

\inst{
Department of Physics, Nara Women's University, Nara 630-8506
}

\recdate{\hspace{3.5cm} }

\abst{ 
Magnetic properties in a charge-ordered state  are examined 
for the extended Hubbard model at half-filling. 
Magnetic excitations, magnetic susceptibilities and a nuclear spin 
relaxation rate are calculated 
with taking account of fluctuations around 
the mean-field solution.   
The relevance of the present results to the observation in the 1:1 
organic conductors, (TTM-TTP)I$_3$, is discussed. 
}

\kword{extended Hubbard model, half-filling, (TTM-TTP)I$_3$, 
charge order, magnetism
}

\begin{document}
\sloppy
\maketitle

Organic conductors are one of the most suitable materials for 
studying exotic electronic correlation effects in low dimensional systems.
Quasi one-dimensional (1-D) quarter-filled conductors, 
Bechgaard salts (TMTSF)$_2$X
and their sulfur analog (TMTTF)$_2$X, have been intensively studied 
and it has been found that 
the rich phases, {\it e.g.}, spin-Peierls states, 
spin density wave (SDW) state, superconductivity and so on, 
are realized as a function of temperature, $T$, and (effective)
pressure.\cite{review}    

Recently, the 1:1 organic material, (TTM-TTP)I$_3$, 
has been synthesized~\cite{Mori-BCSJ} and studied experimentally.\cite{Tajima,Mori-PRL,Maesato,Fujimura,Onuki} 
By extended H$\ddot{\rm u}$ckel molecular orbital calculation, 
the transfer integral along the stacking direction 
is estimated as $t_c$ = 0.26eV and is much larger than the others, 
$t_a$, $t_b$ $<$0.01eV.\cite{Mori-PRL} 
Therefore, the electronic structure is highly 1-D along the stacking axis. 
Different from (TMTSF)$_2$X and (TMTTF)$_2$X, 
the compound has a half-filled band.
Although a 1-D electron system at half-filling is in 
the Mott insulating state, 
the material shows metallic conduction for higher temperature 
and exhibits metal-insulator (MI) transition at $T_c =$ 120-160 K.\cite{Mori-PRL,Maesato} 
Below this temperature region, the gap in the spin excitation has been observed 
in measurement of magnetic susceptibility\cite{Maesato,Fujimura,Onuki} and 
nuclear spin relaxation rate.\cite{Onuki}
In addition,  X-ray
measurements~\cite{Maesato,Fujimura} and NMR spectrum~\cite{Onuki}
indicate the appearance of the charge order below $T_c$. 
Thus, the MI transition of this material 
is followed by the charge order and the spin gap,
and is different from the transition in the 
quasi 1-D quarter-filled organic materials.

Since nearest-neighbor repulsion plays a crucial role 
for the charge order, 
the 1-D extended Hubbard model (EHM) with on-site, $U$, and nearest-neighbor
repulsion, $V$, is considered to be appropriate for studying 
the above properties. 
The ground state of 1-D EHM at half-filling has been investigated 
theoretically.\cite{Bari,Cabib,Hirsch,Cannon-1,Cannon-2,Voit,Dongen,Zhang} 
Recent investigations by numerical\cite{Nakamura,Sengupta} 
and analytical calculations\cite{Tsuchiizu} 
have shown the phase diagram on the plane of $U$ and $V$ as follows. 
In the case of large $U$ and $V$, the SDW state is dominant for $U \gsim
2V$, whereas for $U \lsim 2V$ the charge density wave (CDW) state 
corresponding to the charge order is realized 
where the charge-rich and poor sites appear alternatively.   
The transition between the two kinds of density waves is the first order. 
In the case of small $U$ and $V$, in addition to the two kinds of density
waves,  
the bond charge density wave (BCDW) state appears in the narrow region between 
the SDW and CDW states. 
The transition from BCDW to SDW or CDW is continuous. 
We note that the phase diagram closely resembles  
that by mean-field theory~\cite{Cabib} where the first order transition 
occurs at $U = 2V$ and the SDW (CDW) state is realized for 
$U > 2V$ ($U < 2V$).    
To the best of our knowledge, however,  
the properties for finite temperature corresponding to observation 
in (TTM-TTP)I$_3$ and excitation spectra have not been investigated.     

In the present paper, 
magnetic properties of the charge-ordered state in the 1-D EHM
at half-filling are examined based on the mean-field treatment.
Since the mean-field theory reproduces the phase diagram very well  
except the BCDW state, and the interchain coupling exists 
in the real materials, the treatment based on the mean-field solution
seems to be appropriate for describing the properties 
in (TTM-TTP)I$_3$ except the weak interaction region with $U \simeq 2V$.       
Gaussian fluctuations 
around the mean-field solution are taken into account by 
using the path integral method. 
The magnetic excitation spectra, 
magnetic susceptibilities and a nuclear spin relaxation rate are investigated.  

We consider the 1-D EHM given by the following Hamiltonian, 
${\cal H} = {\cal H}_0 + {\cal H}_{\rm int}$, where 
\begin{eqnarray}
{\cal H}_0 
&=& - t \sum_{j,s} 
   \left( c_{j,s}^\dagger  c_{j+1,s}  + {\rm h.c.}\right) 
= \sum_{k,s} \ek c_{k, s}^\dagger  c_{k,s},  \\ 
{\cal H}_{\rm int} &=& \frac{U}{2} \sum_{j,s} 
n_{j,s} n_{j, -s} + 
V \sum_{j,s,s'} n_{j,s} n_{j+1,s'}  \nonumber \\
&=& \sum_{-Q_0 < q \leq Q_0} 
\Big\{(\frac{U}{4N} + \frac{V}{N} \cos q a) n(q)n(-q) \nonumber \\ 
& & \hspace{5em} - \frac{U}{4N} m(q) m(-q) \Big\}.  
\end{eqnarray}
Here, $t$ is  
  the transfer energy between the nearest-neighbor site, 
 $\epsilon_k = -2t \cos ka$ with lattice constant $a$, and
$Q_0 = \pi /a$. 
The quantity,  
$c_{j,s}^\dagger ( = 1/\sqrt{N} \sum_k \e^{-\im k x_j} 
c_{k,s}^\dagger) $,  denotes the creation
operator of the electron at the $j$-th site with spin $s$, 
$n(q) = \sum_{j,s} \e^{\im q x_j} c^\dagger_{j,s} c_{j,s}$, 
$m(q) = \sum_{j,s} \e^{\im q x_j} s c^\dagger_{j,s} c_{j,s}$, 
and $N$ is the number of lattices (and electrons). 
We express the action corresponding to the above Hamiltonian 
by Grassmann algebras and decompose the interaction parts 
into bilinear forms by utilizing the Stratonovich-Hubbard transformation.
The resulting action is written as,   
\begin{eqnarray}
{\cal S} 
&=& \int_0^{\beta} \d \tau 
\Big\{ \sum_{k,s} c^*_{k,s}(\tau) (\deltau - \mu + \epsilon_k) c_{k,s}(\tau) \nonumber \\
&+& \sum_{-Q_0 < q \leq Q_0} \Big[
-(\frac{U}{4N} + \frac{V}{N}\cos qa) \rho(q,\tau) \rho(-q,\tau) \nonumber \\
&+& (\frac{U}{2N} + \frac{2V}{N}\cos qa) \rho(q,\tau) n(-q,\tau) \nonumber \\
&+& \frac{U}{4N} \sigma(q,\tau) \sigma(-q,\tau) 
- \frac{U}{2N} \sigma(q,\tau) m(-q,\tau) \Big] \Big\}, 
\end{eqnarray}
where $\beta = 1/T$ and $\mu$ is the chemical potential.
Here, we express 
$\nu(0,\tau) = \nu_0 + \delta \nu(0,\tau)$, 
$\nu(Q_0,\tau) = \nu_{Q_0} + \delta \nu(Q_0,\tau)$ and
$\nu(q,\tau) = \delta \nu(q,\tau)$ 
with $0<|q|<Q_0$ ($\nu = \rho$ or $\sigma$), and expand $\cal S$ up to 
the second order of
$\delta \nu$. 
Then the electron degrees of freedom are integrated out.  
The mean-field equations determining $\nu_0$ and $\nu_{Q_0}$ 
are obtained under the condition where
the average of the terms including the first order fluctuation vanish.
In case of $U < 2V$, the charge-ordered state, 
where $\sigma_0 = \sigma_{Q_0} =0$, becomes most stable, and 
the energy gap $\Delta = (U/2 - 2V)\rho_{Q_0}/N$ is determined by  
\begin{eqnarray}
 \Delta &=& \frac{2V - U/2}{N} \sum_k \frac{\Delta}{E_k}
\left\{f(-E_k)-f(E_k)\right\}, 
\end{eqnarray}
with $f(x) = 1/(\e^{\beta x} +1)$ and $E_k = \sqrt{\ek^2 + \Delta^2}$. 
The chemical potential is given by 
$\mu = (U/2 + 2V) \rho_0/N$ with $\rho_0 = N$. 

The effective action up to the second order
is obtained as 
${\cal S}_{\rm eff} = \sum_{\nu = \rho, \sigma}
{\cal S}^{\nu}_{\rm eff}$, 
where 
${\cal S}^{\nu}_{\rm eff} = \sum_{\on, -Q_0 < q \leq Q_0}
A^\nu(q,\im \on) \delta \nu (-q, -\im \on) \delta \nu (q, \im \on) 
$. 
The quantity, $A^\nu(q,\im \on)$, is given as follows, 
\begin{eqnarray}
 A^{\rho}(q,\im \on) &=& - (\frac{U}{4N}+\frac{V}{N}\cos qa) \nonumber \\
&\times&
\left\{1 - (\frac{U}{N}+\frac{4V}{N}\cos qa)K(q,\im \on)\right\}, \\
 A^{\sigma}(q,\im \on) &=& \frac{U}{4N}
\left\{1 + \frac{U}{N}K(q,\im \on)\right\}, 
\end{eqnarray}
where
\begin{eqnarray}
& & K(q, \im \on) = \sum_{-Q_0 < k \leq Q_0} \sum_{\alpha = \pm, \gamma = \pm}
\frac{1}{4}\nonumber \\
&\times& 
(\frac{f(E_k) - f(\alpha E_{k+q})}{E_k - \alpha E_{k+q} + \im \gamma \on}) 
(1 + \alpha \frac{\ek \ekq + \Delta^2}{\Ek \Ekq}).  
\end{eqnarray}
It should be noted that there is no coupling 
between the fluctuation with the wavenumber
$q$ and that with $q-Q_0$ although the lattice periodicity
changes to $2a$ due to formation of the charge order.  
The reason for the absence of the above coupling will be discussed below. 

Charge and spin susceptibilities, $\chi_\rho (q,\im \on)$ 
and $\chi_\sigma (q,\im \on)$, and the local susceptibilities,
$\chi_\nu(x_i,x_i ; \im \on)$,  
are given as follows,
\begin{eqnarray}
 \chi_\rho(q,\im \on) &=& \frac{1}{N} \Big\{ \langle\delta \rho(q,\im \on)
\delta \rho(-q,-\im \on)\rangle \nonumber \\ 
& & + \frac{2N}{U + 4V \cos qa} \Big\}, \\
 \chi_\sigma(q,\im \on) &=& \frac{1}{N}\left\{\langle\delta \sigma(q,\im \on)
\delta \sigma(-q,-\im \on)\rangle - \frac{2N}{U}\right\}, \nonumber \\ 
& & \\
\chi_\nu(x_i, x_i ; \im \on) &=& \frac{1}{N}\sum_q \chi_\nu (q, \im \on), 
\end{eqnarray}
where
\begin{eqnarray}
\langle \delta \nu(q,\im \on) \delta \nu (-q, - \im \on)\rangle 
=
\frac{1}{2 A^\nu(q,\im\on)}. 
\end{eqnarray}
We note that the spatially alternating component 
of the local susceptibilities, 
which is  
written as $(-1)^{i} / {N^2} \sum_q \langle \delta \nu(q,\im \on)
\delta \nu(-q+Q_0,-\im \on)\rangle$,
vanishes
due to $\langle \delta \nu(q,\im \on)
\delta \nu(-q+Q_0,-\im \on)\rangle = 0$, \ie 
the absence of coupling between the fluctuation with $q$ and that
with $q-Q_0$.  
Thus, the local susceptibilities in the charge-rich site are the same
as those in the poor site.  
It is understood in terms of the particle-hole (p-h) transformation as
follows.
The Hamiltonian, eqs.(1) and (2), are invariant under the p-h
transformation, $c_{j,\sigma} \to (-1)^j c_{j,\sigma}^\dagger$.
On the other hand, the amount of charge in the charge-rich site 
changes into that in the poor site and
{\it vice versa} by the p-h transformation. 
Thus, the p-h transformation exchanges 
the charge-rich site for the poor site.    
Therefore, the local quantities at the charge-rich site are equal to 
those at the poor site. 
In the following, we discuss the spin degree of freedom in detail. 

At first, we investigate the spin excitation spectra, which are given by
the solutions of $\chi_{\sigma}(q,\omega)^{-1} = 0$. 
We show $\omega (q)$ at $T = 0$ for $U/t=3.0$ and $V/t=2.0$ in Fig.1
where the solid curve represents the collective mode and the dotted curves, $\omega_{\rm c}^{\rm min}$ and $\omega_{\rm
c}^{\rm max}$, 
express the minimum and maximum values of the continuum spectra,
respectively.
\begin{figure}
\centerline{\epsfxsize=8.0cm\epsfbox{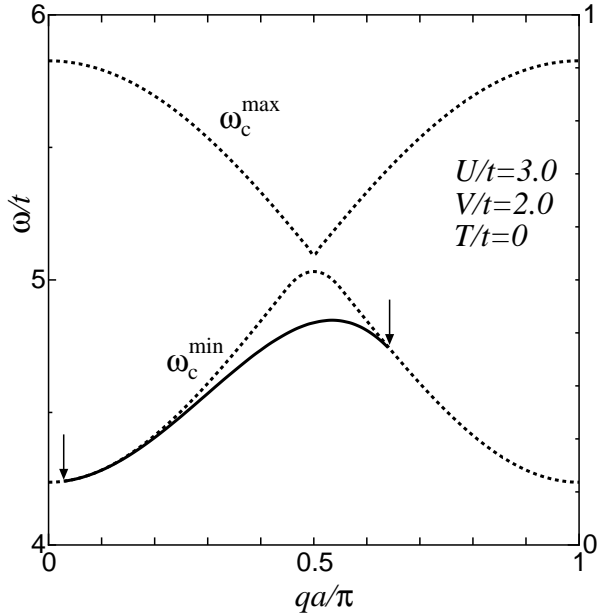}}
\caption{ 
The spin excitation spectra, $\omega$, 
as a function of $q$ at $T/t=0$ in the case of 
$U/t=3.0$ and $V/t=2.0$. 
Here the solid curve represents the collective mode and 
the dotted curves, $\omega_{\rm c}^{\rm min}$ and $\omega_{\rm
c}^{\rm max}$, 
express the minimum and maximum value of the continuum spectra,
 respectively. 
The arrows indicate the locations where the collective mode 
touches the continuum spectra.   
}
\label{fig:band}
\end{figure}
The collective mode appears just below $\omega_{\rm c}^{\rm min}$ and 
has the maximum value for $q = q_{\rm m} \gsim \pi/(2a)$.
In addition, the collective mode goes into 
the continuum spectra at $q \gsim 0$ and $q > q_{\rm m}$.
The locations where the mode touches the continuum spectra are indicated
by the arrows in Fig.1.  

Next we discuss the magnetic susceptibility 
and nuclear spin relaxation rate, $T_1^{-1}$,
given by $(T_1 T)_i^{-1} \propto \lim_{\omega \to 0} 
{\rm Im}\chi_\sigma(x_i, x_i ; \omega)/\omega $. 
As was discussed above,  
both the local spin susceptibility and 
$(T_1 T)_i^{-1}$ are independent of the site. 
These results indicate that 
the split of NMR spectra is not a Knite shift but due to the other
origin such as a chemical shift, and the nuclear spins in crystallographically 
equivalent atoms decay with one kind of relaxation time. 
Note that it is not obvious whether the same conclusions are obtained or not 
for the charge-ordered states coexisting with SDW found in
quasi 1-D quarter-filling organic
conductors.\cite{Pouget-Ravy,Hiraki-Kanoda}
This is because 
the p-h transformation does not exchange 
the amount of charge or spin in the rich site for that
in the poor site
though the Hamiltonian is invariant under the transformation.     
We show the uniform susceptibility, $\chi_\sigma(0,T)$, 
the staggered susceptibility, $\chi_\sigma(Q_0,T)$,
and $R(T) \equiv (T_1 T)^{-1}$ as a function of $T$ in Fig.2. 
Here $\chi_0(0,0)$ and $R_0(0)$ are the susceptibility and $(T_1 T)^{-1}$
at $T=0$ in the absence of interaction.    
\begin{figure}
\centerline{\epsfxsize=8.0cm\epsfbox{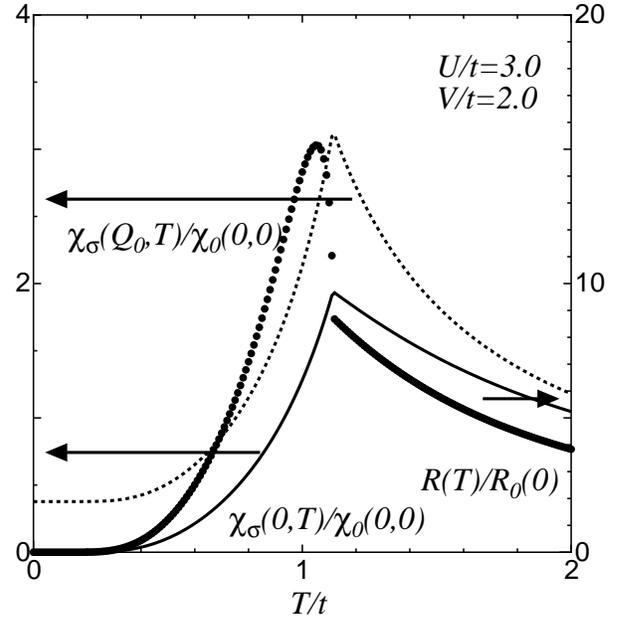}}
\caption{ 
The magnetic susceptibilities, $\chi_\sigma(0,T)$ and
 $\chi_\sigma(Q_0,T)$, 
 normalized by $\chi_0(0,0)$ and $R(T) \equiv (T_1 T)^{-1}$
normalized by $R_0(0)$ 
as a function of $T$  in the case of $U/t=3.0$ and $V/t=2.0$.
Here $\chi_0(0,0)$ and $R_0(0)$ are respectively the susceptibility at
 $q=0$ and $T=0$ 
and $(T_1 T)^{-1}$ at $T=0$ in the absence of the interaction.   
}
\label{fig:T-dep}
\end{figure}
For $T>T_c$ with $T_c$ being the critical temperature of the CDW, 
these quantities increase with decreasing temperature. 
Since $T_c$ is of the order of Fermi energy, $2t$, 
Curie-like behavior in $\chi_\sigma(0,T)$ is observed.  
On the other hand, below $T_c$, the susceptibilities 
decrease rapidly. 
The quantity, $R(T)$, is enhanced just below $T_c$ and shows a rapid decrease 
for $T \ll T_c$. 
The behavior is the same as Hebel-Schlichter peak in $s$-wave
superconductivity.\cite{Gruner} 
We note that both quantities, $\chi_\sigma(Q_0, T)$ and $R(T)$, 
tend to infinity at $T = T_c$ for $U = 2 V$, 
\ie the boundary between the CDW and the SDW states 
because the transition between the two density waves is the first order.  

Here we consider the magnitude of the interactions. 
From the reflectance spectra, 
the on-site interaction, $U$, 
and the hopping integral, $t$, 
are estimated as $U \simeq$ 0.57eV and $t \simeq 0.16$eV at 100K\cite{Tajima}
where the nearest-neighbor repulsion is neglected.  
Since there is a possibility that 
the magnitude of interaction estimated by the optical
measurement is too big for discussing 
the thermodynamic properties,\cite{Note} 
we may expect smaller values of interactions.
In fact, as is shown in Fig.3, 
when we use $U/t = 0.8$ and $V/t = 0.6$, the transition
temperature is close to that for (TTM-TTP)I$_3$ and 
the magnetic susceptibility for $T \gsim T_c$ does not show 
Curie-like behavior
as was observed in the experiments.
In addition, the above choice leads to the resistivity independent of
$T$ for high temperature, which is observed in the
experiments,\cite{Mori-PRL,Maesato}   
because the resistivity at half-filling above the charge gap behaves as
$T^{4K_\rho -3}$,\cite{Giamarchi}
and $K_\rho = \sqrt{\left\{1-V/(2\pi t)\right\}/\left\{1+(U+5V)/(2\pi
t)\right\}} \simeq 3/4$ in this case (see
inset of Fig.3). 
The magnitude of the interactions we have pointed out are in the weak
coupling regime where the BCDW state appears for $U \simeq 2V$. 
However, the estimation is considered to be sensible      
as long as the values of the interactions do not belong to the
BCDW region in the phase diagram determined numerically\cite{Nakamura} and analytically.\cite{Tsuchiizu}    
In fact, the values, $U/t = 0.8$ and $V/t=0.6$, are   
in the CDW region of the phase diagram.   
\begin{figure}
\centerline{\epsfxsize=8.0cm\epsfbox{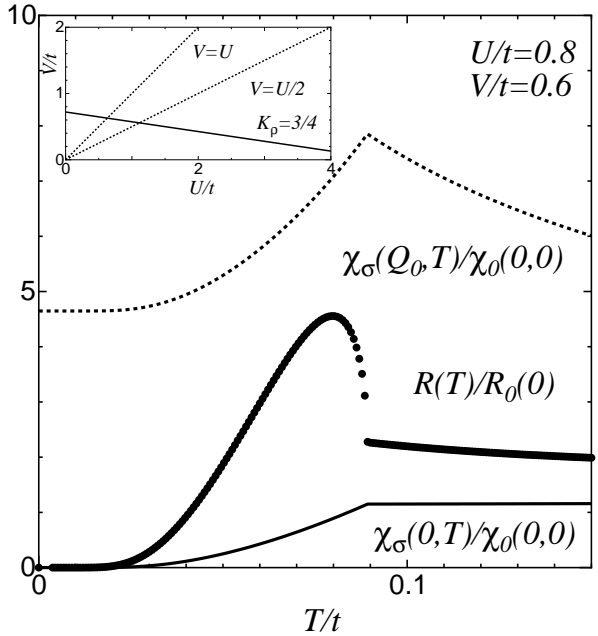}}
\caption{ 
The magnetic susceptibilities, $\chi_\sigma(0,T)$ and $\chi_\sigma(Q_0,T)$,
 normalized by $\chi_0(0,0)$ and $R(T) \equiv (T_1 T)^{-1}$
normalized by $R_0(0)$ 
as a function of $T$  in the case of $U/t=0.8$ and $V/t=0.6$.
Inset : the curve expressing $K_\rho = 3/4$ (solid curve) on the plane of
 $U/t$ and $V/t$, where the upper (lower) dotted line expresses $V=U$ ($V=U/2$). 
}
\label{fig:T-dep-est}
\end{figure}

In conclusion, we investigated the magnetic properties of 1-D EHM 
at half-filling with taking account of fluctuation around the mean-field 
solution. 
The magnetic excitation spectra at $T=0$, 
temperature dependence of  
the magnetic susceptibilities and that of the nuclear spin relaxation
rate are discussed.  
The local magnetic susceptibility and the nuclear spin
relaxation rate are found to be  
independent of the site even in the charge-ordered state. 
In addition, we pointed out the possibility that 
the magnitude of the interaction of the 1:1 material, (TTM-TTP)I$_3$, 
is small compared to that estimated from the 
reflectance spectra.   

\section*{Acknowledgments}
The author would like to thank T. Takahashi, Y. Suzumura, S. Iwabuchi,
K. Kawasaki,  
K. Yonemitsu, H. Kontani, K. Hiraki, M. Tsuchiizu and
 Y. Hashizume for useful discussion.
This work was supported by Grant-in-Aid for Encouragement of 
Young Scientists (No. 13740220), and for Scientific Research (A)
(No. 13304026) and (C) (No. 14540302)
from the Ministry of Education, Culture, Sports, Science and Technology,  
Japan.


\begin{thebibliography}{99}

\bibitem{review}
For review, see,  
T. Ishiguro, K. Yamaji and G. Saito: 
Organic Superconductors (Springer-Verlag, Berlin, 1998).
\bibitem{Mori-BCSJ}
T. Mori, H. Inokuchi, Y. Misaki, T. Yamabe, H. Mori and S. Tanaka: 
\jo{\BCSJ}{67}{1994}{661}.
\bibitem{Tajima}
H. Tajima, M. Arifuku, T. Ohta, T. Mori, Y. Misaki, T. Yamabe H. Mori
	and S. Tanaka: 
\jo{\SM}{71}{1995}{1951}. 
\bibitem{Mori-PRL}
T. Mori, T. Kawamoto, J. Yamaura, T. Enoki, Y. Misaki, T. Yamabe,
 H. Mori and S. Tanaka: 
\jo{\PRL}{79}{1997}{1702}. 
\bibitem{Maesato}
M. Maesato, Y. Sasou, S. Kagoshima, T. Mori, T. Kawamoto, Y. Misaki and
	T. Yamabe:
\jo{\SM}{103}{1999}{2109}.
\bibitem{Fujimura}
N. Fujimura, A. Namba, T. Kambe, Y. Nogami, K. Oshima, T. Mori,
	T. Kawamoto,
Y. Misaki and T. Yamabe:
\jo{\SM}{103}{1999}{2111}.
\bibitem{Onuki}
M. Onuki, K. Hiraki, T. Takahashi, D. Jinbo, T. Kawamoto, T. Mori,
	K. Tanaka and Y. Misaki:
\jo{\JPCS}{62}{2001}{405}.
\bibitem{Bari}
R. A. Bari: 
\jo{\PRB}{3}{1971}{2662}.
\bibitem{Cabib}
D. Cabib and E. Callen:
\jo{\PRB}{12}{1975}{5249}.
\bibitem{Hirsch}
J. E. Hirsch: 
\jo{\PRL}{53}{1984}{2327}.
\bibitem{Cannon-1}
J.W. Cannon and E. Fradkin: 
\jo{\PRB}{41}{1990}{9435}.
\bibitem{Cannon-2}
J.W. Cannon, R.T. Scalettar and E. Fradkin: 
\jo{\PRB}{44}{1991}{5995}.
\bibitem{Voit}
J. Voit: 
\jo{\PRB}{45}{1992}{4027}.
\bibitem{Dongen}
P.G.J.van Dongen:
\jo{\PRB}{49}{1994}{7904}.
\bibitem{Zhang}
G. P. Zhang: 
\jo{\PRB}{56}{1997}{9189}.
\bibitem{Nakamura}
M. Nakamura: 
\jo{\JPSJ}{68}{1999}{3123}, 
\jo{\PRB}{61}{2000}{16377}.
\bibitem{Sengupta}
P. Sengupta, A.W. Sandvik and D.K. Campbell:
\jo{\PRB}{65}{2002}{155113}.
\bibitem{Tsuchiizu}
M. Tsuchiizu and A. Furusaki:
\jo{\PRL}{88}{2002}{056402}.
\bibitem{Pouget-Ravy}
J.P. Pouget and S. Ravy: 
\jo{J.\ Phys.\ I (France)}{6}{1996}{1501}, 
\jo{\SM}{85}{1997}{1532}.
\bibitem{Hiraki-Kanoda}
K. Hiraki and K. Kanoda: 
\jo{\PRL}{80}{1998}{4737}.
\bibitem{Gruner}
The enhancement of $(T_1 T)^{-1}$ just below $T_c$ in the CDW state is 
discussed in the following text,    
G. Gr$\ddot{\rm u}$ner: 
{\it Density Waves in Solids}, Sec.3 (Addison-Wesley, New York, 1994). 
\bibitem{Note}
see P.102 in ref. \citen{review}.
\bibitem{Giamarchi}
T. Giamarchi: 
\jo{\PRB}{44}{1991}{2905}.

\end{thebibliography}
\end{document}